\documentclass[11pt,a4paper]{article}
\pdfoutput = 1
\usepackage{jheppub}
\usepackage[T1]{fontenc} 

\usepackage{graphicx}
\usepackage[space]{grffile}
\usepackage{latexsym}
\usepackage{amsfonts,amsmath,amssymb}
\usepackage{url}
\usepackage[utf8]{inputenc}
\usepackage{mathrsfs}

\DeclareMathAlphabet{\mathpzc}{OT1}{pzc}{m}{it}

\usepackage{natbib}

\title{Direct measurement of $\alpha_{\rm QED}(m_{\rm Z}^2)$ at the FCC-ee}

\author{{\bf{Patrick Janot}}\\{\it {CERN, PH Department, Geneva, Switzerland}}}

\abstract{
When the measurements from the FCC-ee become available, an improved determination of the standard-model "input" parameters will be needed to fully exploit the new precision data towards either constraining or fitting the parameters of beyond-the-standard-model theories. Among these input parameters is the electromagnetic coupling constant estimated at the Z mass scale, $\alpha_{\rm QED}(m^2_{\rm Z})$. The measurement of the muon forward-backward asymmetry at the FCC-ee, just below and just above the Z pole, can be used to make a direct determination of $\alpha_{\rm QED}(m^2_{\rm Z})$ with an accuracy deemed adequate for an optimal use of the FCC-ee precision data.}

\begin{document}
  
\maketitle

\section{Introduction}
\label{sec:Introduction}
  The design study of the Future Circular Colliders (FCC) in a 100-km ring in the Geneva area has started at CERN at the beginning of 2014, as an option for post-LHC particle accelerators. The study has an emphasis on proton-proton and electron-positron high-energy frontier machines~\cite{FCCWebSite}. In the current plans, the first step of the FCC physics programme would exploit a high-luminosity ${\rm e^+e^-}$ collider called FCC-ee, with centre-of-mass energies ranging from the Z pole to the ${\rm t\bar t}$ threshold and beyond. A first look at the physics case of the FCC-ee can be found in Ref.~\cite{Bicer_2014}.

In this first look, an estimate of the achievable precision on a number of Z-pole observables was inferred and used in a global electroweak fit to set constraints on weakly-coupled new physics up to a scale of 100\,TeV~\cite{Ellis_2015}. These constraints were obtained under two assumptions: {\it (i)} the precision of the pertaining theoretical calculations will match the expected experimental accuracy by the time of the FCC-ee startup; and {\it (ii)} the determination of standard-model input parameters -- four masses: $m_{\rm Z}$, $m_{\rm W}$, $m_{\rm top}$, $m_{\rm Higgs}$; and three coupling constants: $\alpha_s(m^2_{\rm Z})$, $G_{\rm F}$, $\alpha_{\rm QED}(m^2_{\rm Z})$ -- will improve in order not to be the limiting factors to the constraining power of the fit. The determinations of the Higgs boson mass from the LHC data~\cite{Aad_2015} and of the Fermi constant from the muon lifetime measurement~\cite{MuLan_2013} are already sufficient for this purpose. It is argued in Refs.~\cite{Bicer_2014,2015arXiv151205194D} that the FCC-ee can adequately improve the determination of the other three masses and of the strong coupling constant by one order of magnitude or more: the experimental precision targets for the FCC-ee are 100\,keV for the Z-boson mass, 500\,keV for the W-boson mass, 10\,MeV for the top-quark mass, and 0.0001 for the strong coupling constant. (The FCC-ee also aims at reducing the Higgs boson mass uncertainty down to 8\,MeV.)

No mention was made, however, of a way to improve the determination of the electromagnetic coupling constant evaluated at the Z mass, and it was simply assumed that a factor $5$ improvement with respect to today's uncertainty -- down to $2\times 10^{-5}$ -- could be achieved by the time of the FCC-ee startup. Today,  $\alpha_{\rm QED}(m^2_{\rm Z})$ is determined from $\alpha_{\rm QED}(0)$ (itself known with an accuracy of $10^{-10}$) with the running coupling constant formula:
\begin{equation}
\alpha_{\rm QED}(m^2_{\rm Z}) = \frac{\alpha_{\rm QED}(0)}{1 - \Delta\alpha_\ell(m^2_{\rm Z}) - \Delta\alpha^{(5)}_{\rm had}(m^2_{\rm Z})}.
\end{equation}
Its uncertainty is dominated by the experimental determination of the hadronic vacuum polarization, $\Delta\alpha^{(5)}_{\rm had}(m^2_{\rm Z})$, obtained from the dispersion integral:
\begin{equation}
\Delta\alpha^{(5)}_{\rm had}(m^2_{\rm Z}) = \frac{\alpha m^2_{\rm Z}}{3\pi} \int_{4m_\pi^2}^\infty \frac{R_\gamma(s)}{s(m^2_{\rm Z}-s)}ds,
\end{equation}
where $R_\gamma(s)$ is the hadronic cross section $\sigma^0({\rm e^+e^-} \to \gamma^\ast \to {\rm hadrons})$ at a given centre-of-mass energy $\sqrt{s}$, normalized to the muon pair cross section at the same centre-of-mass energy. At small values of $\sqrt{s}$, typically up to 5\,GeV, and in the $\Upsilon$ resonance region from 9.6 to 13\,GeV, the evaluation of the dispersion integral relies on the measurements made with low-energy ${\rm e^+e^-}$ data accumulated by the KLOE, CMD-2/SND, BaBar, Belle, CLEO and BES experiments. The most recent re-evaluation~\cite{Davier_2011,Davier_2012} gives $\Delta\alpha^{(5)}_{\rm had}(m^2_{\rm Z}) = (275.7 \pm 1.0)\times 10^{-4}$, which leads to 
\begin{equation}
\label{eq:alphaQED}
\alpha_{\rm QED}^{-1}(m_{\rm Z}^2) = 128.952 \pm 0.014,
\end{equation}
corresponding to a relative uncertainty on the electromagnetic coupling constant, $\Delta\alpha/\alpha$, of $1.1\times 10^{-4}$. It is hoped that future low-energy  ${\rm e^+e^-}$ data collected by the BES~III and VEPP-2000 colliders will improve this figure to $5 \times 10^{-5}$ or better~\cite{Jegerlehner_2011}.

In this study, it is shown that the FCC-ee can provide another way of determining the electromagnetic coupling constant with a similar or better accuracy, from the precise measurement of muon forward-backward asymmetry, $A_{\rm FB}^{\mu\mu}$, just above and just below the Z peak, as part of the resonance scan. This method does not rely on the experimental determination of the vacuum polarization $\Delta\alpha^{(5)}_{\rm had}$. Here, the point is not to extrapolate $\alpha_{\rm QED}(m^2_{\rm Z})$ from $\alpha_{\rm QED}(0)$, but to provide a direct evaluation of $\alpha_{\rm QED}$ at $\sqrt{s} \simeq m_{\rm Z}$, hence with totally different theoretical and experimental uncertainties. This measurement would in turn be combined with other determinations for an even smaller uncertainty.   
  
This letter is organized as follows. In Section~\ref{sec:theory}, the reasons for the choice of $A_{\rm FB}^{\mu\mu}$ as an observable sensitive to $\alpha_{\rm QED}$ are given, and the sensitivity is determined as a function of the centre-of-mass energy. The optimal centre-of-mass energies, as well as the integrated luminosities and running time needed to achieve a statistical uncertainty of a few $10^{-5}$ are determined in Section~\ref{sec:stat}. Possible systematic uncertainties are discussed and evaluated in Section~\ref{sec:syst}. 

\section{The muon forward-backward asymmetry and the electromagnetic coupling constant}
\label{sec:theory}
  
\begin{figure}[htbp]
\begin{center}
\includegraphics[width=0.7\columnwidth]{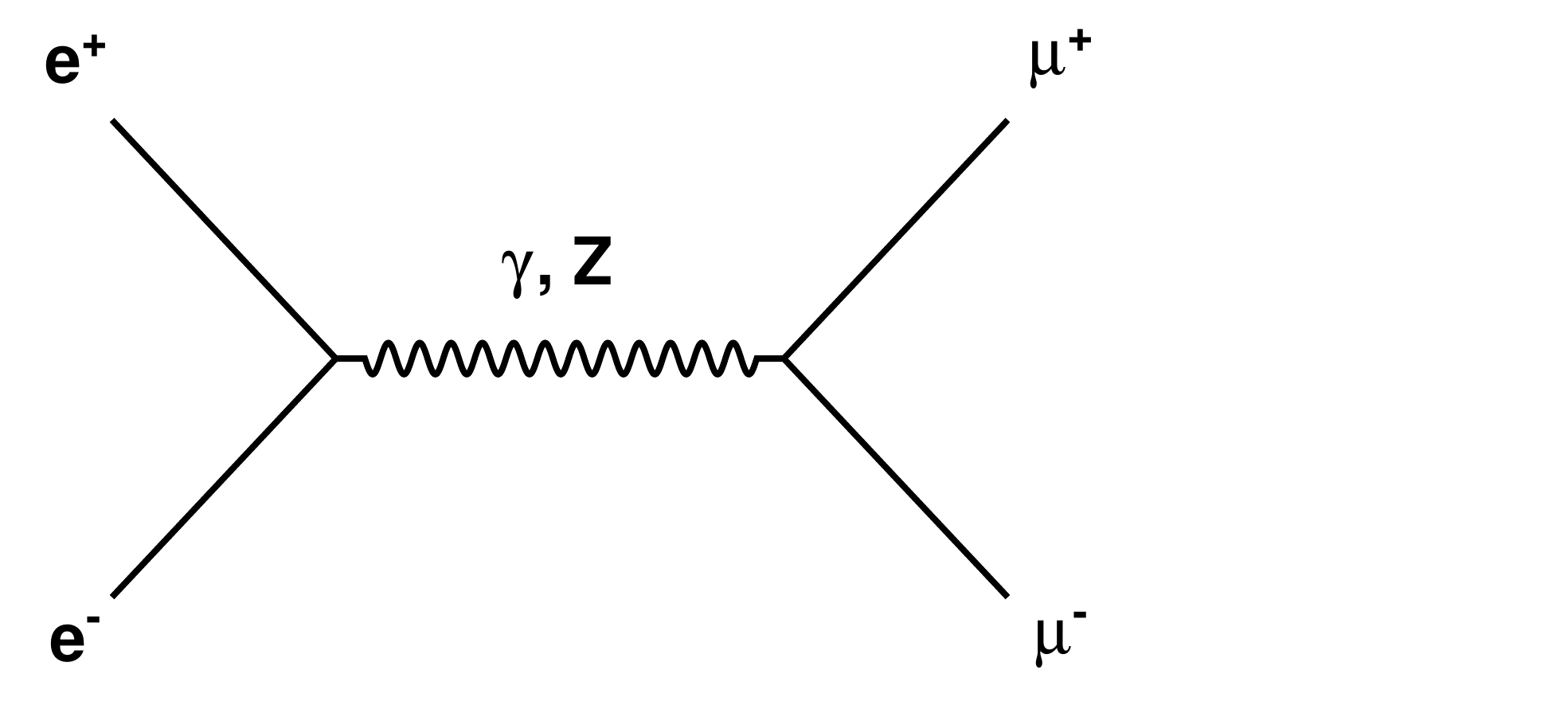}
\caption{\label{fig:graph}
Tree-level Feynmann graph for $\mu^+\mu^-$ production at the FCC-ee%
}
\end{center}
\end{figure}

At the FCC-ee, the muon pair production proceeds via the graph depicted in Fig.~\ref{fig:graph} through either a Z or a $\gamma$ exchange. 

  At tree level, the cross section $\sigma_{\mu\mu}$ therefore contains three terms: {\it (i)} the $\gamma$-exchange term squared, proportional to $\alpha^2_{\rm QED}(s)$; {\it (ii)} the Z-exchange term squared, proportional to $G_{\rm F}^2$ (where $G_{\rm F}$ is the Fermi constant); and {\it (iii)} the $\gamma$-Z interference term, proportional to $\alpha_{\rm QED}(s) \times G_{\rm F}$. These three terms are denoted ${\cal G}$, ${\cal Z}$, and ${\cal I}$ in the following. Their expressions as a function of the centre-of-mass energy $\sqrt{s}$ can be found in Ref.~\cite{Leike_1991} and reported below.
\begin{eqnarray}
{\cal G} & = & \frac{\displaystyle c_\gamma^2}{\displaystyle s}, \\
{\cal Z} & = & \frac{\displaystyle c^2_{\rm Z} ({\mathpzc v^2 + a^2})^2 \times s}{\displaystyle (s-m_{\rm Z}^2)^2+m_{\rm Z}^2\Gamma_{\rm Z}^2}, \\
{\cal I} & = & \frac{\displaystyle 2c_\gamma c_{\rm Z} {\mathpzc v^2} \times (s-m_{\rm Z}^2)}{\displaystyle (s-m_{\rm Z}^2)^2+m_{\rm Z}^2\Gamma_{\rm Z}^2},
\end{eqnarray}
with the following definitions:
\begin{equation}
c_\gamma = \sqrt{\frac{4\pi}{3}} \alpha_{\rm QED}(s), \ \ c_{\rm Z} = \sqrt{\frac{4\pi}{3}} \frac{m^2_{\rm Z}}{2\pi} \frac{G_{\rm F}}{\sqrt{2}}, \ \ {\mathpzc a} = -\frac{1}{2},\ \ {\mathpzc v} = {\mathpzc a} \times ( 1 - 4\sin^2\theta_{\rm W}),
\end{equation}
and where $\theta_{\rm W}$ is the effective Weinberg angle ($\sin^2\theta_{\rm W} \simeq 0.2315$).

\begin{figure}[htbp]
\begin{center}
\includegraphics[width=0.95\columnwidth]{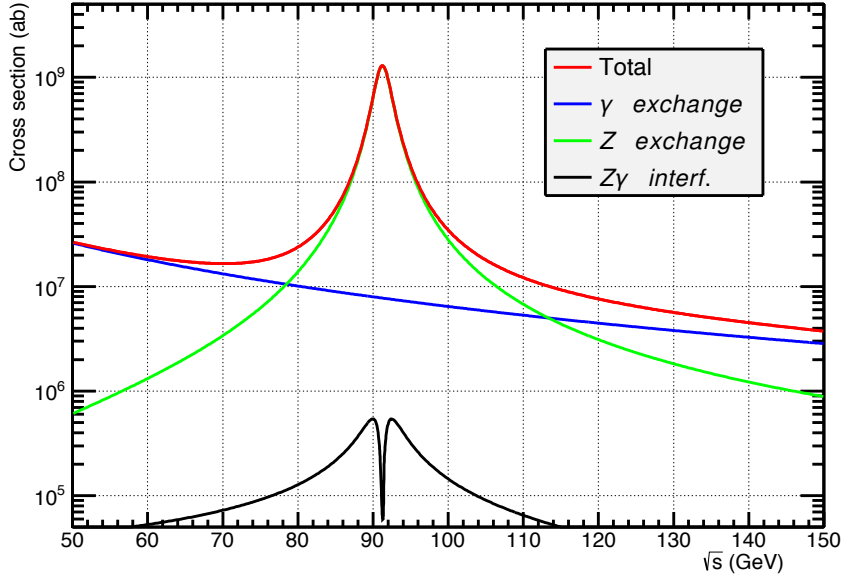}
\caption{\label{fig:mumucross}
Cross section for the ${\rm e^+e^-} \to \mu^+\mu^-$ process (red curve) and the three contributions, calculated from the analytical expressions of Ref.\cite{Leike_1991}: pure $\gamma$-exchange term (blue curve); pure Z-exchange term (green curve); and the absolute value of the $\gamma$-Z interference term (black curve). The initial-state radiation is included, and $s^\prime/s$ is required to exceed 0.99.}
\end{center}
\end{figure}

An absolute measurement of the $\mu^+\mu^-$ production cross section $\sigma_{\mu\mu} = {\cal Z} + {\cal I} + {\cal G}$ is therefore a priori sensitive to $\alpha_{\rm QED}$ through the interference term and the $\gamma$-exchange term. The cross section and the three contributing terms are displayed in Fig.~\ref{fig:mumucross} as a function of the centre-of-mass energy $\sqrt{s}$, with the inclusion of initial state radiation (ISR). In this figure, the effective collision energy after ISR, denoted $\sqrt{s^\prime}$, is required to satisfy $s^\prime \geqslant 0.99 s$. The importance of such a requirement on $s^\prime$, together with the way to control it experimentally, is discussed in Section~\ref{sec:ISR}.

At a given $\sqrt{s}$, a small variation $\Delta\alpha$ of the electromagnetic coupling constant translates to a variation $\Delta\sigma_{\mu\mu}$ of the cross section :
\begin{equation}
\Delta\sigma_{\mu\mu} = \frac{\Delta\alpha}{\alpha}({\cal I} + 2{\cal G}).
\end{equation}
As is well visible in Fig.~\ref{fig:mumucross}, the interference term can be neglected in the above equation. As a consequence, if the cross section can be measured with a precision $\Delta\sigma_{\mu\mu}$, the relative precision on the electromagnetic coupling constant amounts to  
\begin{equation}
\frac{\Delta\alpha}{\alpha} \simeq \frac{\Delta\sigma_{\mu\mu}}{2{\cal G}} \simeq \frac{1}{2} \frac{\Delta\sigma_{\mu\mu}}{\sigma_{\mu\mu}} \left( 1 + \frac{\cal Z}{\cal G} \right).
\end{equation}

The target statistical precision of $2\times10^{-5}$ on $\alpha_{\rm QED}$ can therefore be achieved with more than $10^9$ $\mu^+\mu^-$ events and at centre-of-mass energy where the Z contribution to the cross section is much smaller than the photon contribution. These two conditions call for a centre-of-mass energy smaller than 70\,GeV, where the cross section is both large and dominated by the photon contribution. Beside the fact that this centre-of-mass energy is not in the current core programme of the FCC-ee and that the needed integrated luminosity of $50\,{\rm ab}^{-1}$ would require at least a year of running at this energy in the most favourable conditions, the measurement itself poses a number of intrinsic difficulties. Indeed, the absolute measurement of a cross section with a precision of a few $10^{-5}$ requires the selection efficiency, the detector acceptance, and the integrated luminosity to be known with this precision or better. Even if not impossible to meet, these requirements are exceedingly challenging in the extraction of $\alpha_{\rm QED}$ from this method with the needed precision.  

The muon forward-backward asymmetry, $A_{\rm FB}^{\mu\mu}$, defined as 
\begin{equation}
A_{\rm FB}^{\mu\mu} = \frac{\sigma_{\mu\mu}^{\rm F} - \sigma_{\mu\mu}^{\rm B}}{\sigma_{\mu\mu}^{\rm F} + \sigma_{\mu\mu}^{\rm B}},
\end{equation}
where $\sigma_{\mu\mu}^{\rm F (B)}$ is the $\mu^+\mu^-$ cross section for events with the $\mu^-$ direction in the forward (backward) hemisphere with respect to the ${\rm e}^-$-beam direction, hence with $\sigma_{\mu\mu}^{\rm F} + \sigma_{\mu\mu}^{\rm B} = \sigma_{\mu\mu}$, solves most of these obstacles. Indeed, it is a self-normalized quantity, which thus does not need the measurement of the integrated luminosity. Moreover, most uncertainties on the selection efficiency and the detector acceptance simply cancel in the ratio. This observable is therefore a good candidate for a measurement with an exquisite precision.

At lowest order, and if the terms proportional to $m_\mu^2/m_{\rm Z}^2 \sim 10^{-6}$ are neglected, the angular distribution of the $\mu^-$ from the ${\rm e^+e^-} \to \mu^+\mu^-$ production can be written in the following way~\cite{Altarelli:1989hv}:
\begin{equation}
\label{eq:generalAFB}
\frac{d\sigma_{\mu\mu}}{d\cos\theta}(s) \propto G_1(s) \times (1+\cos^2\theta) + G_3(s) \times 2\cos\theta,
\end{equation}
where $G_1(s)$ and $G_3(s)$ can be expressed as a function of $\cal G$, $\cal Z$ and $\cal I$ as follows: 
\begin{equation}
\label{eq:G1G3}
G_1(s) = {\cal G} + {\cal I} + {\cal Z} {\rm \ \ \  and \ \ \ } G_3(s) = \frac{{\mathpzc a^2}}{{\mathpzc v^2}}\left\{ {\cal I} + \frac{4 {\mathpzc v^4/a^4}}{\left(1+{\mathpzc v^2/a^2}\right)^2}{\cal Z} \right\}.
\end{equation}
After integration over the muon polar angle $\theta$, the forward-backward asymmetry therefore amounts to:
\begin{equation}
\label{eq:asymumu}
A_{\rm FB}^{\mu\mu}(s) = \frac{3}{4} \frac{G_3(s)}{G_1(s)}.
\end{equation}
  
\begin{figure}[htbp]
\begin{center}
\includegraphics[width=0.93\columnwidth]{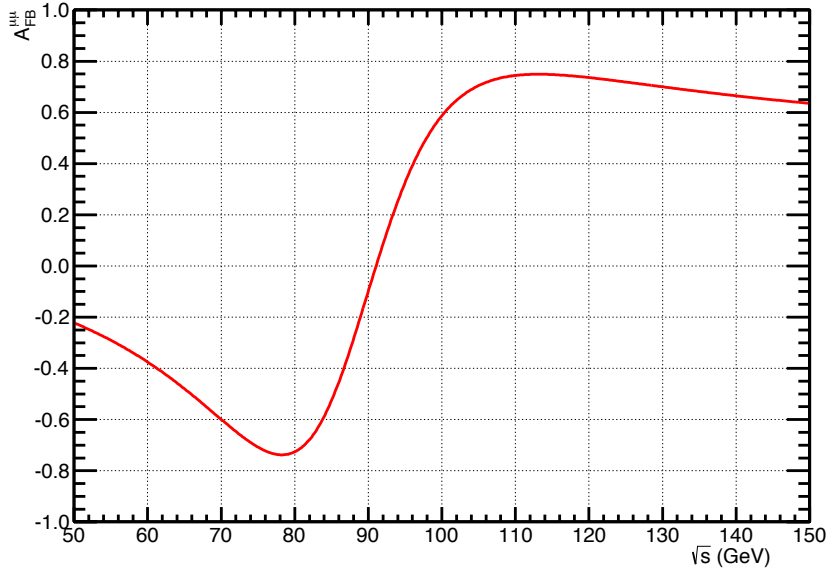}
\caption{\label{fig:asymumu}
The muon forward-backward asymmetry in ${\rm e^+e^-} \to \mu^+\mu^-$ as a function of the centre-of-mass energy.%
}
\end{center}
\end{figure}

The variation of $A_{\rm FB}^{\mu\mu}$ as a function of the centre-of-mass energy, as obtained from Eq.~\ref{eq:asymumu}, is shown in Fig.~\ref{fig:asymumu}. In the above expressions, the photon-exchange term is totally symmetric, hence is absent from the numerator. Because ${\mathpzc v^4/a^4} \simeq 3\times 10^{-5}$, the Z-exchange term contribution to the asymmetry is minute, except at the Z pole where the interference term vanishes and the asymmetry is small: $A_{\rm FB,0}^{\mu\mu} = (3/4) \times 4{\mathpzc v^2 a^2}/({\mathpzc a^2+v^2})^2 \simeq 0.016$. The interference term, on the other hand, is almost 100\% anti-symmetric and contributes mostly to the numerator. (The contribution of the interference term to the denominator, {\it i.e.}, to the total cross section, can be neglected as shown in Fig.~\ref{fig:mumucross}.) 

\begin{figure}[htbp]
\begin{center}
\includegraphics[width=0.95\columnwidth]{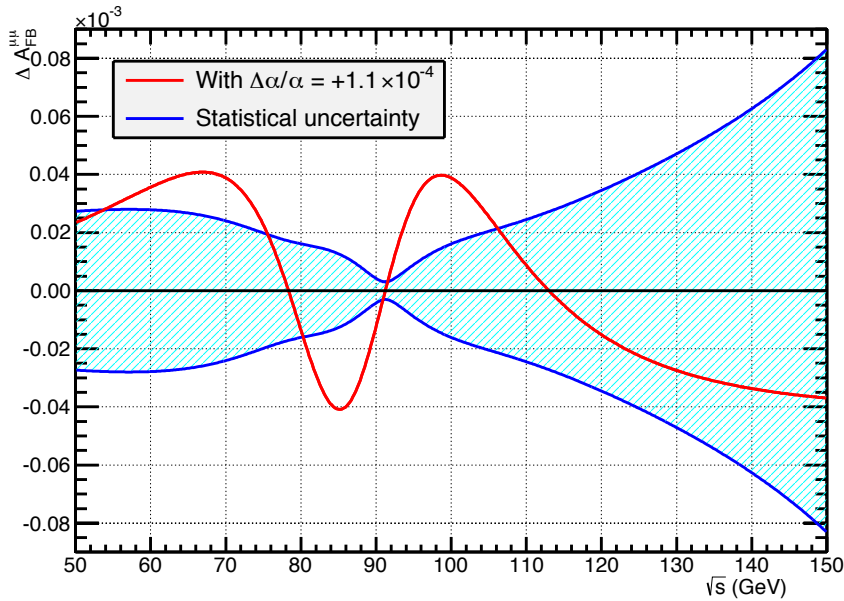}
\caption{\label{fig:sensitivity}
The red curve shows the variation of the muon forward-backward asymmetry as a function of $\sqrt{s}$ for a relative change of $\alpha_{\rm QED}(s)$ by $+1.1 \times 10^{-4}$. The asymmetry has no sensitivity to $\alpha_{\rm QED}$ when the red curve crosses the black horizontal line. The blue area represents the absolute statisitical uncertainty with which the muon forward-backward asymmetry can be measured at the FCC-ee in one year of data taking at any given centre-of-mass energy.%
}
\end{center}
\end{figure}

The off-peak muon forward-backward asymmetry can therefore be expressed as follows:
\begin{equation}
\label{eq:AFBsimple}
A_{\rm FB}^{\mu\mu} = A_{\rm FB,0}^{\mu\mu} + \frac{3}{4} \frac{{\mathpzc a^2}}{{\mathpzc v^2}} \frac{\cal I}{{\cal G}+{\cal Z}}.
\end{equation}
At a given $\sqrt{s}$, a small variation $\Delta\alpha$ of the electromagnetic coupling constant translates to a variation $\Delta A_{FB}^{\mu\mu}$ of the muon forward-backward asymmetry:
\begin{equation}
\Delta A_{\rm FB}^{\mu\mu} = \frac{\Delta\alpha}{\alpha} \times \frac{3}{4} \frac{{\mathpzc a^2}}{{\mathpzc v^2}} \frac{{\cal I} ({\cal Z}-{\cal G})}{({\cal G}+{\cal Z})^2} = \left( A_{\rm FB}^{\mu\mu} - A_{\rm FB,0}^{\mu\mu} \right) \times \frac{{\cal Z}-{\cal G}}{{\cal Z}+{\cal G}} \times \frac{\Delta\alpha}{\alpha}.
\end{equation}
In first approximation, the asymmetry is therefore not sensitive to $\alpha_{\rm QED}$ when the Z- and photon-exchange terms are equal, {\it i.e.}, at $\sqrt{s} = 78$ and $112$\,GeV (Fig.~\ref{fig:mumucross}), where the asymmetry is maximal (Fig.~\ref{fig:asymumu}). Similarly, the sensitivity to the electromagnetic coupling constant vanishes in the immediate vicinity of the Z pole. The red curve of Fig.~\ref{fig:sensitivity} shows the variation of $A_{\rm FB}^{\mu\mu}$ for a relative change of $\alpha_{\rm QED}$ by $+1.1 \times 10^{-4}$, as a function of $\sqrt{s}$. In other words, the red curves displays the absolute precision with which $A_{\rm FB}^{\mu\mu}$ must be measured to start improving the accuracy on $\alpha_{\rm QED}(m^2_{\rm Z})$ with respect to today's determination. 

For a positive variation of $\Delta\alpha$, the sign of $\Delta A_{\rm FB}^{\mu\mu}$, {\it i.e.}, the sign of  $\left( A_{\rm FB}^{\mu\mu} - A_{\rm FB,0}^{\mu\mu} \right) \times ({\cal Z}-{\cal G})$, changes at each of these centre-of mass energies: it is positive below 78\,GeV, where the asymmetry is negative and the Z contribution is smaller than the photon contribution, becomes negative between 78\,GeV and the Z pole, where the Z contribution dominates, then positive again from the Z pole all the way to 112\,GeV because the asymmetry becomes positive, and negative for larger centre-of-mass energies where the photon contribution takes over. This interesting property, in particular the sign change around the Z pole, is fully exploited in Section~\ref{sec:syst}.  Written the other way around and in a perhaps more useful manner for the following, the relative precision on the electromagnetic coupling constant amounts to  
\begin{equation}
\label{eq:dalphaoveralpha}
\frac{\Delta\alpha}{\alpha} = \frac{\Delta A_{\rm FB}^{\mu\mu} }{A_{\rm FB}^{\mu\mu}- A_{\rm FB,0}^{\mu\mu}} \times \frac{{\cal Z}+{\cal G}}{{\cal Z}-{\cal G}} \simeq \frac{\Delta A_{\rm FB}^{\mu\mu}}{A_{\rm FB}^{\mu\mu}} \times \frac{{\cal Z}+{\cal G}}{{\cal Z}-{\cal G}},
\end{equation}
where the approximation in the last term of the equality is valid off the Z peak. 

\section{Statistical power of the method}
\label{sec:stat}
  
\begin{figure}[htbp]
\begin{center}
\includegraphics[width=0.94\columnwidth]{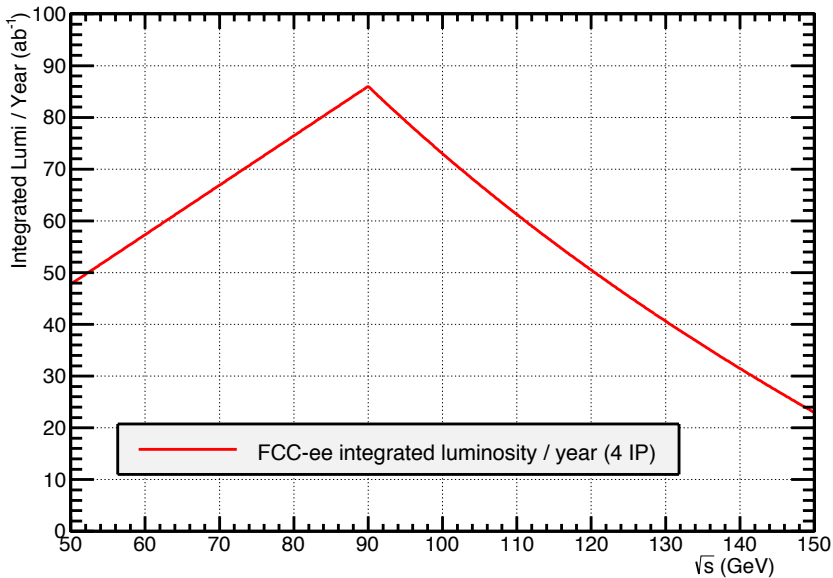}
\caption{\label{fig:lumi}
Target integrated luminosities for the FCC-ee, in a scheme with four interaction points, for centre-of-mass energies between 50 and 150\,GeV.%
}
\end{center}
\end{figure}

 The optimal centre-of-mass energies are those which minimize the statistical uncertainty on $\alpha_{\rm QED}(s)$. For a given integrated luminosity ${\cal L}$, the statistical uncertainty on the forward-backward asymmetry amounts to 
\begin{equation}
\label{eq:statAFB}
\sigma \left( A_{\rm FB}^{\mu\mu} \right)= \sqrt{\frac{1-{A_{\rm FB}^{\mu\mu}}^2}{\cal L \sigma_{\mu\mu}}}.
\end{equation}
The target luminosities for the FCC-ee in a configuration with four interaction points are $215 \times 10^{34} {\rm cm}^{-2}{\rm s}^{-1}$ per interaction point at the Z pole and $38 \times 10^{34} {\rm cm}^{-2}{\rm s}^{-1}$ per interaction point at the WW pair production threshold~\cite{Zimmermann_2015}. With $10^7$ effective seconds per year, the total integrated luminosity is therefore expected to be $86~{\rm ab}^{-1}$/\,year at the Z pole and $15.2~{\rm ab}^{-1}$/\,year at the WW threshold. Between these two points, the variation of the luminosity with the centre-of-mass energy is assumed to follow a simple power law: ${\cal L}(\sqrt{s}) = {\cal L}(m_{\rm Z}) \times s^a$. The very large Z pole luminosity is achieved by colliding about 60,000 bunches of electrons and positrons, which fill the entirety of the 400\,MHz RF buckets available over 100\,km. It also corresponds to a time between two bunch crossings of 5\,ns, which is close to the minimum value acceptable today for the experiments. With a constant number of bunches, the luminosity was therefore conservatively assumed to linearly decrease with the centre-of-mass energy (and reach $0.$ for $\sqrt{s} = 0.$), leading to the profile of Fig.~\ref{fig:lumi}.      

With the cross section of Fig.~\ref{fig:mumucross}, the asymmetry of Fig.~\ref{fig:asymumu}, and the integrated luminosity of Fig.~\ref{fig:lumi}, Eq.~\ref{eq:statAFB} leads to the statistical uncertainty on $A_{\rm FB}^{\mu\mu}$ displayed as the blue area in Fig.~\ref{fig:sensitivity}, for a one-year running at any given centre-of-mass energy. An improvement on the determination of $\alpha_{\rm QED}(s)$ is possible wherever the red curve lies outside the blue area, and is largest when the absolute value of the ratio between the red and blue curves is maximum. 

\begin{figure}[htbp]
\begin{center}
\includegraphics[width=0.94\columnwidth]{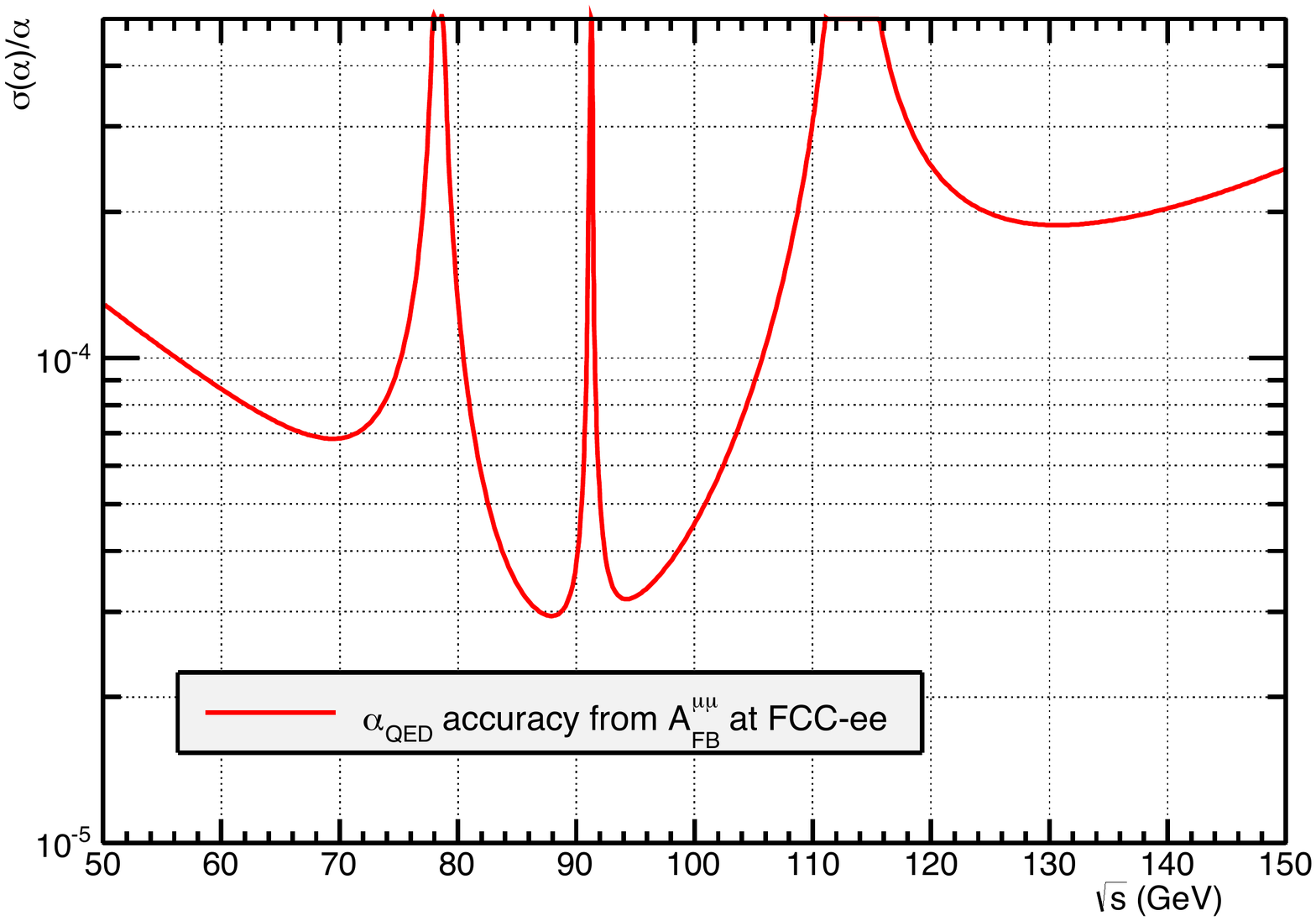}
\caption{\label{fig:alphaQED}
Relative statistical uncertainty for the $\alpha_{\rm QED}(s)$ determination from a measurement of the muon forward-backward asymmetry at the FCC-ee, with a one-year running at any given centre-of-mass energy. The best accuracy is obtained for one year of running either just below or just above the Z pole.%
}
\end{center}
\end{figure}

The corresponding relative accuracy for the $\alpha_{\rm QED}(s)$ determination is shown in Fig.~\ref{fig:alphaQED}. The best accuracy of $\sim 3\times 10^{-5}$ is obtained for one year of running either just below or just above the Z pole, specifically at $\sqrt{s_-} \sim 87.9$\,GeV and $\sqrt{s_+} \sim 94.3$\,GeV. 

The value of the electromagnetic coupling constant extracted from the muon forward-backward asymmetry measured at either energy, $\alpha_- \equiv \alpha_{\rm QED}(s_-)$ and $\alpha_+ \equiv \alpha_{\rm QED}(s_+)$, are then extrapolated towards a determination of $\alpha_0 \equiv \alpha_{\rm QED}(m_{\rm Z}^2)$ with the running coupling constant expression around the Z pole, valid at all orders in the leading-log approximation:
\begin{equation}
\label{eq:betafunction}
\frac{1}{\alpha_0} = \frac{1}{\alpha_\pm} + \beta\log\frac{s_\pm}{m_{\rm Z}^2}, 
\end{equation}
where $\beta$ is proportional to the well-known QED $\beta$-function. In the standard model and at the lowest QED/QCD order, it reads $\beta_0 = \sum_{\rm f} Q_{\rm f}^2/3\pi$, where the sum runs over all active fermions at the Z pole (f = e, $\mu$, $\tau$, d, u, s, c b) and $Q_{\rm f}$ is the fermion electric charge in unit of $e$. The standard model extrapolation correction from $\alpha_\pm$ to $\alpha_0$ therefore amounts to $-0.033$ from the measurement below the Z pole, and $+0.030$ from the measurement above the Z pole, corresponding to a relative correction of $\pm 2.5 \times 10^{-4}$ in both cases, {\it i.e.}, an order of magnitude larger than the targeted uncertainty on $\alpha_0$. While this correction is known with an excellent precision in the standard model -- the QED $\beta$-function is now known with QED corrections up to five loops and QCD corrections up to four loops~\cite{Kataev:2012rf,Baikov_2012} --, it is certainly preferable to remove this model dependence (and the residual theory uncertainty) from the determination of $\alpha_0$. 

The dual measurements of $\alpha_-$ and of $\alpha_+$ solve this issue and yields the straightforward combination:
\begin{equation}
\label{eq:xi}
\frac{1}{\alpha_0} = \frac{1}{2} \left( \frac{ 1 - \xi}{\alpha_-} + \frac{1+\xi}{\alpha_+} \right), {\rm \ \ \ where \ \ \ } \xi = \frac{\log s_-s_+ / m_{\rm Z}^4 }{\log s_- / s_+ } \simeq 0.045, 
\end{equation}
without any model dependence related to the running of the electromagnetic constant. This combination of a measurement below the Z peak and a measurement above the Z peak has other advantages, the most important of which is the cancellation to a large extent of many systematic uncertainties, as explained in the next section. With this weighted average, the targeted precision of $2\times 10^{-5}$ can be obtained from one year at $87.9$\,GeV and one year at $94.3$\,GeV with the sole measurement of the muon forward-backward asymmetry. The running time at each energy can be reduced to six months -- as is assumed in the following -- if additional measurements are considered, {\it e.g.}, the tau forward-backward asymmetry and, possibly, the electron forward-backward asymmetry.

\section{Systematic uncertainties}
\label{sec:syst}

For the forward-backward asymmetry measurement to be relevant in the determination of the electromagnetic coupling constant, all possible systematic uncertainties must be kept well below the statistical uncertainty aimed at in Section~\ref{sec:stat}. Systematic uncertainties may be of experimental, parametric, and theoretical origin, and are studied in turn below. 

\subsection{Experimental uncertainties}
\label{sec:exptal}

\subsubsection{Beam energy calibration}
\label{sec:beamenergy}

Circular ${\rm e^+e^-}$ colliders have the unique feature of providing the possibility to measure the beam energy with the resonant depolarization method~\cite{Assmann_1999} with an outstanding accuracy. At the FCC-ee~\cite{Koratzinos:2015hya}, this accuracy has been estimated~\cite{Bicer_2014,Mike_2015} to be of the order of 50\,keV around the Z pole, of which 45 (23)\,keV are (un)correlated between all energy points, corresponding to a total relative uncertainty of $10^{-6}$. The derivative of the muon forward-backward asymmetry with respect to the centre-of-mass energy, however, is largest around the Z pole, as can be seen from Fig.~\ref{fig:asymumu}. It is therefore important to check that this expected precision is indeed sufficient. 

At $\sqrt{s_-}$ and $\sqrt{s_+}$, the photon contribution is only 5\% of the total cross section and varies slowly with the centre-of-mass energy around the Z pole (Fig.~\ref{fig:mumucross}): this contribution can be considered as a second order effect in the uncertainty evaluation. Equations~\ref{eq:AFBsimple} and~\ref{eq:dalphaoveralpha} therefore simplify to
\begin{equation}
\label{eq:AFBverysimple}
A_{\rm FB}^{\mu\mu}(s_\pm) \simeq \frac{3}{4} \frac{{\mathpzc a^2}}{{\mathpzc v^2}} \frac{{\cal I}}{{\cal Z}} {\rm \ \ \ and \ \ \ } \frac{\Delta\alpha_\pm}{\alpha_\pm} \simeq \frac{\Delta A_{\rm FB}^{\mu\mu}}{A_{\rm FB}^{\mu\mu}}.
\end{equation}
The dependence of $\cal I$ and $\cal Z$ on $s$ and $m_{\rm Z}$ is given at the beginning of Section~\ref{sec:theory}. The forward-backward asymmetry dependence on $s$ and $m_{\rm Z}$ in the vicinity of the Z pole is simply 
\begin{equation}
\label{eq:AFBsmz}
A_{\rm FB}^{\mu\mu}(s,m_{\rm Z}) \propto (s-m_{\rm Z}^2)/(s m^2_{\rm Z}).
\end{equation}

The uncertainties on $\sqrt{s}$ and $m_{\rm Z}$ both amount to 95\,keV, are dominated by the uncertainty of the beam energy measurement, and are largely correlated as indicated above. The uncorrelated variables are therefore the difference $D = \sqrt{s} - m_{\rm Z}$ and the average $\Sigma = (\sqrt{s} + m_{\rm Z})/2.$, with uncertainties of $\sigma_D = 46$\,keV and $\sigma_\Sigma = 94$\,keV, respectively. A straightforward error propagation yields 
\begin{equation}
\frac{\sigma(A_{\rm FB}^{\mu\mu})}{A_{\rm FB}^{\mu\mu}} \simeq \frac{1}{\sqrt{s}m_{\rm Z}}
\sqrt{ \left(s+m^2_{\rm Z}-\sqrt{s}m_{\rm Z}\right)^2 \frac{\sigma_D^2}{D^2} + \left(s+m^2_{\rm Z}+\sqrt{s}m_{\rm Z}\right)^2 \frac{\sigma_\Sigma^2}{\Sigma^2} },
\end{equation}
which in turn simplifies to, at $\sqrt{s_\pm}$, 
\begin{equation}
\frac{\sigma(\alpha_\pm)}{\alpha_\pm} \simeq \frac{\sigma_{D_\pm}} {D_\pm}, {\rm \ \ \  with \ \ \ } D_\pm = \sqrt{s_\pm} - m_{\rm Z},
\end{equation}
after neglecting the much smaller term proportional to $(\sigma_\Sigma/\Sigma)^2$. Numerically, the relative uncertainties on $\alpha_\pm$, or equivalently on $1/\alpha_\pm$, arising from the beam energy measurement both amount to $1.4\times 10^{-5}$ and are uncorrelated. The uncertainty on the coefficient $\xi$ ($\pm 0.00001$) was found to have a totally negligible contribution ($\pm 2 \times 10^{-9}$) to the relative uncertainty on $\alpha_0$ . Only the (uncorrelated) errors on $\alpha_-$ and $\alpha_+$ contribute. As a consequence, the relative uncertainty on $\alpha_{\rm QED}(m^2_{\rm Z})$ arising from the beam energy measurement amounts to 
\begin{equation}
\frac{\sigma(\alpha_0)}{\alpha_0} \simeq \frac{1}{2} \sqrt{(1-\xi)^2\frac{\sigma^2(\alpha_-)}{\alpha_-^2} + (1+\xi)^2\frac{\sigma^2(\alpha_+)}{\alpha_+^2}} \simeq 1 \times 10^{-5}. 
\end{equation}

\subsubsection{Beam energy spread}
  
At the FCC-ee, the relative beam energy spread $\delta$ for centre-of-mass energies around the Z pole is expected~\cite{Zimmermann_2015} to be of the order of 0.12\%, {\it i.e.}, two orders of magnitude larger than the accuracy of the (average) beam energy measurement. The relative centre-of-mass energy spread $\delta$ is $\sqrt{2}$ times smaller, {\it i.e.}, of the order of 0.08\%. The shift $\Delta A_{\rm FB}^{\mu\mu}$ between the predicted asymmetry and its measured value at a centre-of-mass energy $\sqrt{s_\pm}$ is therefore
\begin{equation}
\Delta A_{\rm FB}^{\mu\mu}(s_\pm) = \frac{1}{\sqrt{2\pi s_\pm} \delta} \int A_{\rm FB}^{\mu\mu}(s) \exp -\frac{\left( \sqrt{s}-\sqrt{s_\pm} \right)^2}{2s_\pm\delta^2} d\sqrt{s} - A_{\rm FB}^{\mu\mu}(s_\pm),
\end{equation}
which yields, with the functional form of $A_{\rm FB}^{\mu\mu}(s)$ given in Eq.~\ref{eq:AFBsmz} expanded around $s_\pm$:
\begin{equation}
\frac{\Delta A_{\rm FB}^{\mu\mu}}{A_{\rm FB}^{\mu\mu}}(s_\pm) \simeq \frac{3 m_{\rm Z}^2}{s_\pm-m_{\rm Z}^2} \delta^2,
\end{equation} 
{\it i.e.}, numerically 
\begin{equation} \frac{\Delta A_{\rm FB}}{A_{\rm FB}}(s_-) = -3.0 \times 10^{-5} {\rm \ \ \  and \ \ \ } \frac{\Delta A_{\rm FB}}{A_{\rm FB}}(s_+) = +3.1 \times 10^{-5},
\end{equation}
under the reasonable assumption that the beam energy spread values are similar at $\sqrt{s_\pm }$ and $m_{\rm Z}$.

The relative changes of $A_{\rm FB}(s_\pm)$ are of the order of the statistical uncertainty, and larger than the uncertainty originating from the beam energy measurement. These changes are, however, of opposite sign, and lead to a remarkable cancellation by more than one order of magnitude in the determination of $\alpha_0$. Indeed, the combination of Eqs.~\ref{eq:dalphaoveralpha} and~\ref{eq:xi} leads to the following estimate of the bias on $\alpha_0$: 
\begin{equation}
\frac{\Delta\alpha_0}{\alpha_0} \simeq 0.528 \frac{\Delta A_{\rm FB}}{A_{\rm FB}}(s_-) + 0.563 \frac{\Delta A_{\rm FB}}{A_{\rm FB}}(s_+) \simeq +1.6 \times 10^{-6}. 
\label{eq:remarkable}
\end{equation}
The uncertainty on this small bias (which is to be corrected for) depends on the accuracy with which the beam energy spread in known. For example, the measurement of bunch length from the distribution of the $\mu^+\mu^-$ event primary vertices determined directly by the FCC-ee experiments would allow a precise determination of the beam energy spread. A precision of 2.5\% could be reached with this method at LEP~\cite{Lancon_1996}, yielding a negligible uncertainty on the $\alpha_{\rm QED}(m^2_{\rm Z})$ determination.

\subsubsection{Muon identification efficiency and detector acceptance}

In Eq.~\ref{eq:asymumu}, the asymmetry is determined under the assumption of a 100\% muon identification efficiency and a $4\pi$ detector acceptance. This equation is still valid for a smaller efficiency, with the condition that it is independent of the muon polar angle. If instead the identification efficiency times the detector acceptance is a non-trivial function of the polar angle, $\varepsilon(\cos\theta)$, the measured muon angular distribution gets modified accordingly, and so does the measured forward-backward asymmetry.

This issue can be solved experimentally with the observation~\cite{Tenchini_2008} that a ${\rm e^+e^-} \to \mu^+\mu^-$ event contains not only a negative muon but also a positive muon, the measured angular distributions of which are given by Eqs.~\ref{eq:generalAFB} and~\ref{eq:asymumu} modified with $\varepsilon(\cos\theta)$: 
\begin{equation}
\frac{{\rm d}N^\pm}{{\rm d}\cos\theta} \propto  \left\{ 1 + \cos^2\theta \pm \frac{8}{3} A_{\rm FB}^{\mu\mu} \cos\theta \right\} \times \varepsilon(\cos\theta),
\end{equation}
where $\varepsilon(\cos\theta)$ is assumed to be  independent on the muon electric charge. This very reasonable assumption can be verified with an adequate accuracy from the  $3\times 10^{11}$ ${\rm Z} \to \mu^+\mu^-$ events collected at $\sqrt{s}= 91.2$\,GeV, by tagging one of the two muons in each event, and probing the other to determine $\varepsilon(\cos\theta)$ separately for positive and negative muons. The ratio of the difference to the sum of the numbers of negative and positive muons detected a given $\cos\theta$ bin, $N_-(\cos\theta)$ and $N_+(\cos\theta)$, therefore amounts to
\begin{equation}
\frac{N_-(\cos\theta) - N_+(\cos\theta)}{N_-(\cos\theta) + N_+(\cos\theta)} = \frac{4}{3}  \frac{2\cos\theta}{1 + \cos^2\theta} A_{\rm FB}^{\mu\mu}, 
\end{equation}
which allows the muon forward-backward asymmetry to be determined in each bin as follows:
\begin{equation}
A_{\rm FB}^{\mu\mu} = \frac{3}{4} \times \frac{N_-(\cos\theta) - N_+(\cos\theta)}{N_-(\cos\theta) + N_+(\cos\theta)} \times \frac{1 + \cos^2\theta}{2\cos\theta},
\end{equation}
an expression from which $\varepsilon(\cos\theta)$ has simplified away in the ratio, hence without any impact on the measurement uncertainty. The muon forward-backward asymmetry for the complete event sample is then obtained by the statistically-weighted average of the bin-by-bin determination over the detector acceptance. Any systematic effect related to the choice of the bin size -- related for example to the muon angular resolution -- can be eliminated by the use of an unbinned likelihood instead.    

\subsubsection{Charge inversion}

The asymmetry of the sample of events where both muon charges are wrongly measured equals $-A_{\rm FB}^{\mu\mu}$. The relative change of the asymmetry arising from double charge inversion therefore amounts to
\begin{equation}
\frac{\Delta A_{\rm FB}^{\mu\mu}}{A_{\rm FB}^{\mu\mu}} = -f^2_\pm,
\end{equation}
where $f_\pm$ is the probability for a muon to be measured with the wrong charge sign. For this effect to be relevant ({\it i.e.}, larger than $2\times 10^{-5}$), $f_\pm$ would need to exceed 0.5\% -- a typical value for LEP detectors. If FCC-ee detectors were similar to LEP detectors, $f_\pm$ would be measured with an outstanding precision from the several $10^7$ million $\mu^\pm\mu^\pm$ events collected at $\sqrt{s_\pm}$, thus allowing the effect to be corrected with no additional uncertainty on the asymmetry. On the other hand, the next generation of detectors for FCC-ee is likely to provide a track momentum resolution better than that of the LEP detectors by up to one order of magnitude, reducing $f_\pm$ to ridiculously small values, with no sizable effect on the asymmetry.  

\subsubsection{Background from tau-pair production}
  
The background from ${\rm e^+e^-}\to \tau^+\tau^-$, where the two taus decay into $\mu\nu_{\rm e}{\nu_\mu}$ has a cross section of the order of 3\% of the ${\rm e^+e^-} \to \mu^+\mu^-$ cross section. It can be greatly reduced by cuts on the muon impact parameters, on the angle between the two muons, and on the muon momenta. The very small residual contribution from this process is however not an issue, as the tau forward-backward asymmetry is expected to be identical to the muon forward-backward asymmetry. No additional uncertainty is therefore expected from this source.

\subsection{Parametric uncertainties}
  
The cross section given in Section~\ref{sec:theory} depends solely on four parameters -- beside $\alpha_{\rm QED}(s)$ -- namley the Fermi constant $G_{\rm F}$, the Z boson mass and width, $m_{\rm Z}$ and $\Gamma_{\rm Z}$, and the Weinberg angle, $\sin^2\theta_{\rm W}$. The precision with which these parameters are known is the source of additional uncertainties for the muon forward-backward asymmetry, and in turn, on the electromagnetic coupling constant. These uncertainties are examined in turn below.

\subsubsection{Z mass}

The uncertainty on the Z mass is fully correlated to the uncertainty on the beam energy. Its effect on the forward backward asymmetry is already taken into account in Section~\ref{sec:beamenergy}.

\subsubsection{Z width}

The Z width simplifies away from the ratio given in Eq.~\ref{eq:AFBverysimple}, which contains only the dominant contributions from ${\cal I}$ and ${\cal Z}$ to the asymmetry. To exhibit the sub-leading dependence on $\Gamma_{\rm Z}$, it is necessary to go back to the more complete expression given in Eq.~\ref{eq:AFBsimple}, which contains also the $\cal G$ term. The uncertainty arising from the knowledge of $\Gamma_{\rm Z}$ is therefore not expected to be dominant. Straightforward error propagation yields
\begin{equation}
\frac{\sigma(A_{\rm FB}^{\mu\mu})}{A_{\rm FB}^{\mu\mu}} = 2 \frac{\cal G}{\cal Z} \times \frac{m_{\rm Z}^2 \Gamma_{\rm Z}^2}{(s-m_{\rm Z}^2)^2+m_{\rm Z}^2 \Gamma_{\rm Z}^2} \times \frac{\sigma_{\Gamma_{\rm Z}}}{\Gamma_{\rm Z}}.
\end{equation}
The uncertainty on $\Gamma_{\rm Z}$ is dominated by the energy calibration error and amounts to $\sigma_D = 46$\,keV, {\it i.e.,} to  about $2\times 10^{-5} \Gamma_{\rm Z}$. At $\sqrt{s_\pm}$, the photon contribution $\cal G$ is about 5\% of the Z contribution $\cal Z$, itself about 50\% of its pole value. As a consequence, the relative uncertainties on $\alpha_\pm$ are equal and amount to $10^{-6}$ with a 100\% correlation. The relative uncertainty on $\alpha_{\rm QED}(m_{\rm Z}^2)$ arising from the Z width is therefore at the same level of $10^{-6}$.
 
\subsubsection{Weinberg angle}

Only the terms proportional to ${\cal I}$ and ${\cal Z}$ in the complete asymmetry expression (Eqs.~\ref{eq:G1G3} and~\ref{eq:asymumu}) vary with $\sin^2\theta_{\rm W}$, through the vectorial coupling $\mathpzc v$. In the vicinity of the Z pole, the small photon contribution can anyway be neglected, and the asymmetry expression simplifies to 
\begin{equation}
\label{eq:ultimateTaylor}
A_{\rm FB}^{\mu\mu} (s) = \frac{3{\mathpzc v^2 a^2}}{({\mathpzc v^2 + a^2})^2} + \frac{c_\gamma}{c_{\rm Z}}\frac{s-m_{\rm Z}^2}{2s}\frac{3{\mathpzc a^2}}{({\mathpzc v^2 + a^2})^2}. 
\end{equation}
The derivative of $A_{\rm FB}^{\mu\mu} (s)$ with respect to $\sin^2\theta_{\rm W}$ can be obtained analytically, yielding in a straightforward manner
\begin{equation}
\frac{\Delta A_{\rm FB}^{\mu\mu}}{A_{\rm FB}^{\mu\mu}}(s) = \frac{\mathpzc 8av}{\mathpzc v^2+a^2} \times
\frac{{\mathpzc a^2-v^2} - \frac{\displaystyle c_\gamma}{\displaystyle c_{\rm Z}}\frac{\displaystyle s-m_{\rm Z}^2}{\displaystyle s}}{{\mathpzc v^2} + \frac{\displaystyle c_\gamma}{\displaystyle c_{\rm Z}}\frac{\displaystyle s-m_{\rm Z}^2}{\displaystyle 2s}} \times \Delta\sin^2\theta_{\rm W}, 
\end{equation}
{\it i.e.}, numerically for $s = s_\pm$:
\begin{equation}
\frac{\Delta A_{\rm FB}^{\mu\mu}}{A_{\rm FB}^{\mu\mu}}(s_-) \simeq -6.92 \Delta \sin^2\theta_{\rm W} 
{\rm \ \ \ and \ \ \ }
\frac{ \Delta A_{\rm FB}^{\mu\mu}}{A_{\rm FB}^{\mu\mu}}(s_+) \simeq +4.87 \Delta \sin^2\theta_{\rm W}.
\end{equation}
The propagation to $\alpha_0$ from Eq.~\ref{eq:remarkable} leads to a partial cancellation by almost one order of magnitude:
\begin{equation}
\frac{\Delta\alpha_0}{\alpha_0} \simeq 0.528 \frac{\Delta A_{\rm FB}}{A_{\rm FB}}(s_-) + 0.563 \frac{\Delta A_{\rm FB}}{A_{\rm FB}}(s_+) \simeq -0.91 \Delta \sin^2\theta_{\rm W}. 
\end{equation}

For the current precision of the effective Weinberg angle determination, of the order $1.6\times 10^{-4}$, this uncertainty on $\alpha_{\rm QED}$ is large and amounts to $1.4 \times 10^{-4}$. At the FCC-ee, however, the measurement of the asymmetry at the Z pole (insensitive to the electromagnetic coupling constant) can be used to improve the precision of the effective Weinberg angle~\cite{Blondel_2015} by a factor 30 to $6 \times 10^{-6}$ -- an uncertainty dominated by the absolute calibration of the beam energy -- thus reducing the uncertainty on $\alpha_{\rm QED}$ to $5 \times 10^{-6}$.  

\subsubsection{Fermi constant}

The Fermi constant is known to a relative accuracy of $5 \times 10^{-7}$~\cite{Olive_2014}, turning into a relative uncertainty on $\alpha_{\rm QED}(m^2_{\rm Z})$ of $5 \times 10^{-7}$.

\subsection{Theoretical uncertainties}
  
Theoretical uncertainties on the muon forward-backward asymmetry arise from the lack of higher orders in the calculation of the muon angular distribution. The dominant higher-order effects on the muon angles originate from QED corrections, namely {\it (i)} initial-state radiation (ISR), {\it i.e.}, the emission of one or several photons by the incoming beams; {\it (ii)} final state radiation (FSR), {\it i.e.}, the emission of photons from the outgoing muons; and {\it (iii)} the interference between ISR and FSR (IFI), which becomes significant when the muons are produced at small angles with respect to the beam axis. The effect of these three QED corrections on the muon angular distributions and on the muon forward-backward asymmetry have been studied analytically in Ref.~\cite{Bardin_1991} with a complete ${\cal O}(\alpha)$ calculation and soft-photon exponentiation, and in a more pragmatic way by the OPAL experiment in Ref.~\cite{Abbiendi_2001}. Their conclusions are summarized and the relevant effects of the ${\cal O}(\alpha^2)$ corrections are examined in this section. Other electroweak corrections are discussed at the end of the section. 

\subsubsection{Final-state radiation (FSR)}
  
Final-state radiation is mostly collinear and is symmetric around the muon directions, at all orders in $\alpha$. The effect on the forward-backward asymmetry is therefore expected to be unmeasurably small~\cite{Abbiendi_2001}. It was checked in Ref.~\cite{Bardin_1991} that the effect is rigorously 0 at order $\alpha$ (with soft-photon exponentiation) if no cut is applied on the final-state photon energy, and vanishingly small if a tight upper cut is applied to the final-state photon energy, typically of the order of $\frac{\alpha}{\pi} \frac{E_{\rm cut}}{\sqrt{s}}$, {\it i.e.},  $\sim 2 \times 10^{-6}$ for $E_{\rm cut} \sim 100$\,MeV. The theoretical uncertainty on this effect due to the ${\cal O}(\alpha^2)$ corrections is typically one-to-two orders of magnitude smaller than that. It is therefore ignored in the following. 

\subsubsection{Initial-state radiation (ISR)}
\label{sec:ISR}
  
Initial-state radiation corrections are known up to order ${\cal O}(\alpha^2)$ with soft-photon exponentiation~\cite{Berends_1988}. Unlike FSR, ISR  has a macroscopic influence on the forward-backward asymmetry. Photons from ISR are emitted mostly along the beam axis, with a twofold consequence: {\it (i)} the centre-of-mass frame of the muon pair therefore acquires a longitudinal boost, which modifies the angular distribution of both muons in a non trivial way; and {\it (ii)} the effective centre-of-mass energy of the collision is reduced to $\sqrt{s^\prime}$ where $s^\prime = (1-x_-)(1-x_+)s$ and $x_\pm = E^\gamma_\pm/\sqrt{s}$ are the fractional radiated energies by the ${\rm e^\pm}$ beams, which modifies the asymmetric term of the cross section through $A_{\rm FB}^{\mu\mu}(s^\prime)$. As $A_{\rm FB}^{\mu\mu}(s^\prime)$ varies quite fast with $\sqrt{s^\prime}$, as displayed in Fig.~\ref{fig:asymumu} and expressed in Eq.~\ref{eq:ultimateTaylor}, a large, negative, variation of the measured asymmetry is indeed to be expected. 
  
When only one ISR photon is radiated by one of the two beams, the effects can be largely mitigated. In the vast majority of the cases, the photon is radiated exactly along the beam axis. The polar angles of the outgoing muons, denoted $\theta^\pm$, suffice in that case to determine the effective centre-of-mass energy $\sqrt{s^\prime}$:
\begin{equation}
\label{eq:sprime}
\frac{s^\prime}{s} = \frac{\sin\theta^+ + \sin\theta^- - \left\vert \sin(\theta^+ +\theta^-)\right \vert }{\sin\theta^+ + \sin\theta^- + \left\vert \sin(\theta^+ +\theta^-) \right\vert}, 
\end{equation}
as well as the $\mu^+$ direction in the centre-of-mass frame of the muon pair
\begin{equation}
\label{eq:cstar}
\cos\theta^\ast = \frac{\sin(\theta^+ - \theta^-)}{\sin\theta^+ + \sin\theta^-}. 
\end{equation}
In this simplest configuration, the use of $\cos\theta^\ast$ entirely corrects for the effect of the longitudinal boost on the angular distribution, and the forward-backward asymmetry dependence on $s^\prime/s$ can be studied explicitly. Furthermore, the events relevant for the determination of $\alpha_\pm$ can be selected by requiring $s^\prime/s$ to be close to unity. If the initial-state photon is radiated with a finite angle with respect to the beam axis, however, Eqs.~\ref{eq:sprime} and~\ref{eq:cstar} no longer hold, but the corresponding events can be rejected by requiring the two muons to be back-to-back in the plane transverse to the beam axis. 

\begin{figure}[htbp]
\begin{center}
\includegraphics[width=0.49\columnwidth]{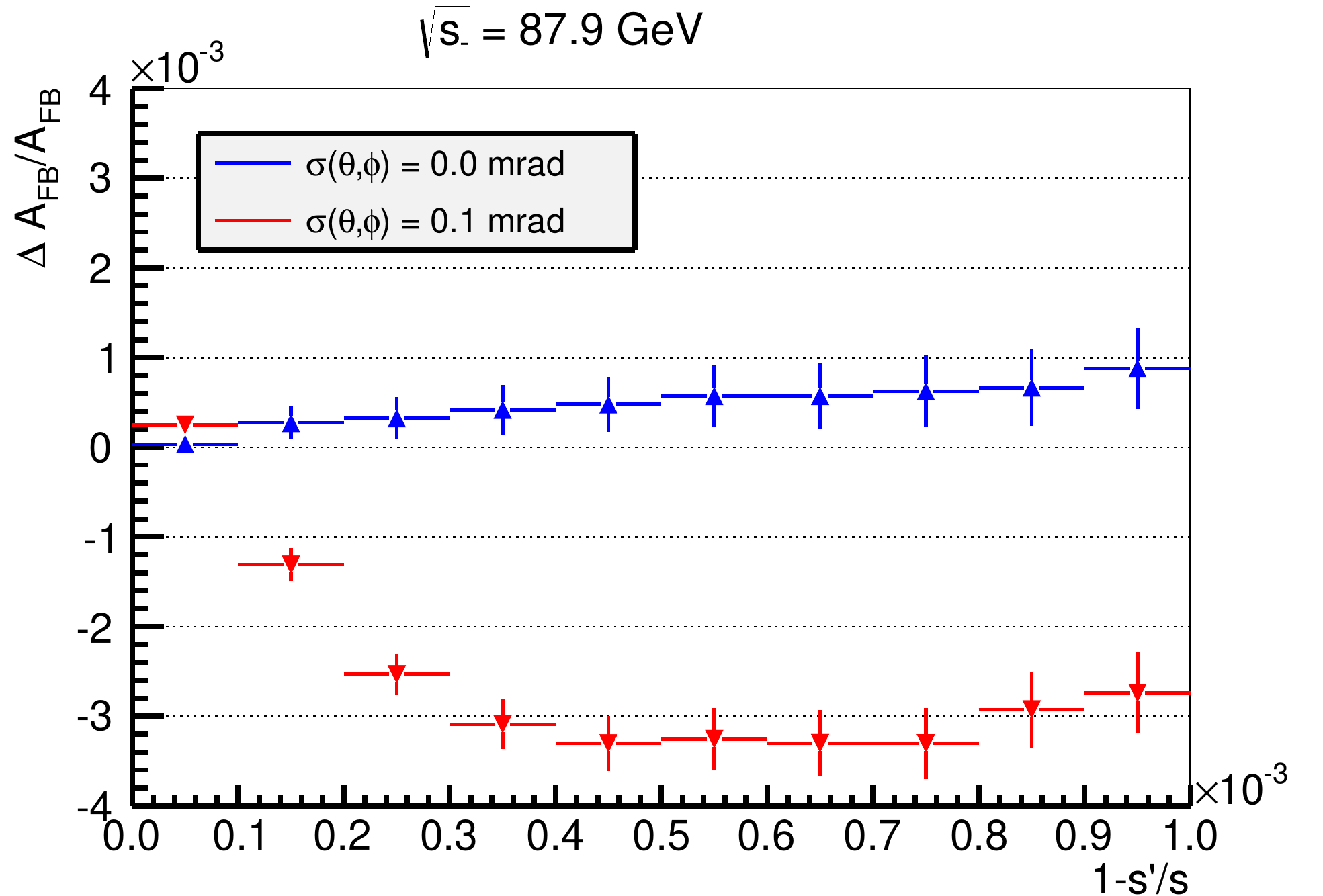}
\includegraphics[width=0.49\columnwidth]{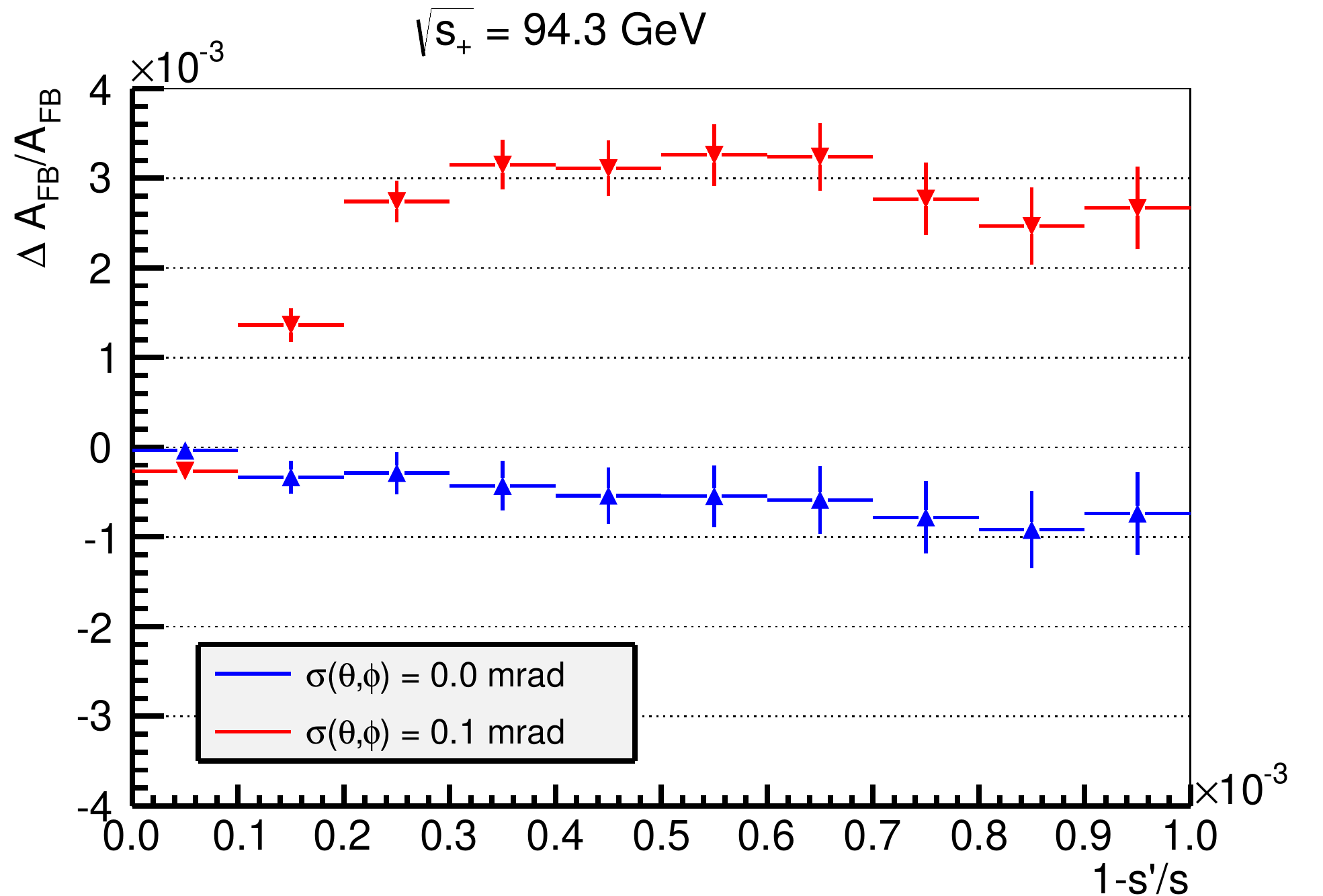}
\end{center}
\caption{\label{fig:ISReffect} Relative bias and statistical uncertainty on the muon forward-backward asymmetry with respect to the standard-model expectation as a function $1-s^\prime/s$ (after the cut $s^\prime/s > 0.999$ is applied) for $\sqrt{s}=\sqrt{s_-}$ (left) and $\sqrt{s_+}$ (right). The blue histogram is obtained with a perfect muon angular resolution, while $\sigma_\theta$ and $\sigma_\phi$ are assumed to amount to 0.1\,mrad in the red histogram. In both cases, $s^\prime/s$ is obtained from the measured angles with Eq.~\ref{eq:sprime}.%
}
\end{figure}

In rare cases, both beams can radiate photons, which render these two equations only approximate, and may still create a bias in the forward-backward asymmetry. To determine the effect of this approximation, large samples of $\mu^+\mu^-$ events were generated at $\sqrt{s_\pm}$. The simulation of ISR was performed with the {\tt REMT} package~\cite{Berends_1985} modified to include ${\cal O}(\alpha^2)$ correcctions with soft-photon exponentiation, and the possibility to radiate up to two photons. For the reasons just explained, only events with $s^\prime/s$ in excess of $0.999$ and an acoplanarity angle between the two muons smaller than 0.35\,mrad were considered. These two cuts typically select about 80\% of the cross section in Fig.~\ref{fig:mumucross}, and tremendously increase the purity towards events without ISR. The blue histograms in Figs.~\ref{fig:ISReffect} show, for $\sqrt{s} = \sqrt{s_-}$ and $\sqrt{s_+}$, the relative biasses on $A_{\rm FB}^{\mu\mu}(s^\prime_\pm)$ with respect to the standard model prediction, as a function of $1-s^\prime/s$ and for a perfect muon angular resolution, $\sigma_\theta = \sigma_\phi = 0$. 

Events with only one ISR photon would lead to a blue straight line at $\Delta A_{\rm FB}/A_{\rm FB} \equiv 0.0$, as $s^\prime/s$ can be exactly determined in that case from Eq.~\ref{eq:sprime}. The possibility to radiate photons from the two beams, however, induces a visible systematic effect on the measured asymmetry, up to 0.1\%, as soon as the energy of these photons is non-zero (i.e., in all but the first bin of the blue histogram). The sign of this effect can be understood as follows: the emission of two photons in opposite directions reduces the effective boost of the $\mu^+\mu^-$ pair, causing the value of $s^\prime$, as determined from Eq.~\ref{eq:sprime}, to be larger than the true effective centre-of-mass energy. The corresponding event migration towards the left of the blue histogram therefore tends to reduce the value of the forward-backward asymmetry (hence a relative increase below the Z peak and a relative decrease above). The effect in the first bin is much smaller than in the other bins because it contains the vast majority of the events, hence is little affected by the migration from bins with a much smaller number of events.   

A non-perfect muon angular resolution also affects the determination of $s^\prime/s$ from Eq.~\ref{eq:sprime} for all events (with or without ISR), hence has a non-trivial effect on the measured asymmetry, as shown in the red histograms of Figs.~\ref{fig:ISReffect} for a uniform angular resolution $\sigma_\theta = \sigma_\phi = 0.1$\,mrad. Because a worse angular resolution would lead to a proportionally larger systematic bias, the measurement of $\alpha_{\rm QED}$ can be used as a benchmark to define the muon reconstruction performance of the FCC-ee detectors. (As an example, the track angular resolution of the pixel detector of SiD~\cite{2013arXiv1306.6329B}, with a hit resolution of $5\,\mu{\rm m}$, a thickness of $1.5\% $ of a radiation length, a radius of 6\,cm, and a length of 34\,cm is better than 0.1\,mrad over the whole acceptance.) For most of the events, without significant ISR photons, the measured angle between the two muons can only decrease from its maximal value of $\pi$ rad, yielding a smaller value of $s^\prime$ from Eq.~\ref{eq:sprime}. Part of these events therefore migrate to the other bins of the distribution, and therefore increase significantly the value of the measured forward-backward asymmetry in these bins, by up to 0.4\% (with a relative variation negative below the Z peak, and positive above). The residual migration from events with initial state radiation, but with two wrongly measured back-to-back muons, has the opposite effect in the first bin of the distribution, albeit with a much smaller amplitude, because of the much larger number of events originally in this bin. 

These two systematic biasses are much larger, by two orders of magnitude, than the target precision with which the forward-backward asymmetry needs to be measured. These biasses can be corrected for if {\it (i)} the energy and angular distributions of initial-state radiation can be predicted with an accuracy better than 1\%, which is probably the case already today; and if {\it (ii)} the muon angular resolution can be mapped with a precision of a few per mil over the whole detector acceptance, which is probably feasible with the large samples of $K^0_{\rm s}$'s, $\Lambda$'s and even $J/\psi$'s expected at the FCC-ee. 

\begin{figure}[htbp]
\begin{center}
\includegraphics[width=0.94\columnwidth]{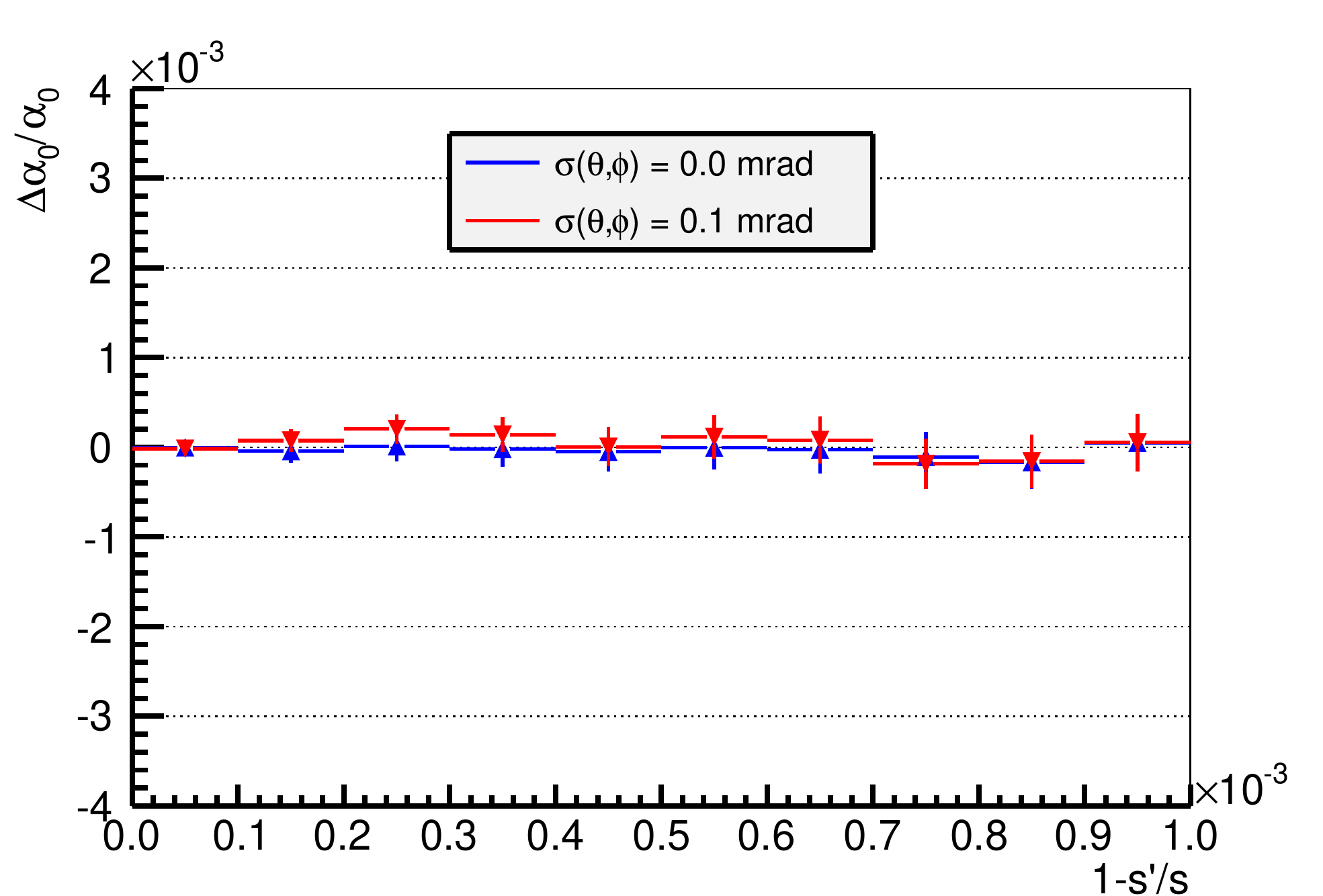}
\caption{\label{fig:ISReffectAll} Relative bias and statistical uncertainty on the electromagnetic coupling constant estimated at the Z mass scale, as a function $1-s^\prime/s$ , from measurements of the muon forward-backward asymmetry at $\sqrt{s_\pm}$, with a perfect muon angular resolution (blue histogram) and with $\sigma_\theta = \sigma_\phi = 0.1$\,mrad (red histogram).%
}
\end{center}
\end{figure}

The predicted relative biasses, however, appear to be quasi-"universal", in the sense that they are similar in amplitude below and above the Z peak, albeit in opposite directions. The combination of the two measurements towards a determination of $\alpha_{\rm QED}(m^2_{\rm Z})$ with Eq.~\ref{eq:remarkable} exhibits an almost perfect cancellation in all bins, as displayed in Fig.~\ref{fig:ISReffectAll} as a function of $1-s^\prime/s$, with the same vertical scale as in Figs.~\ref{fig:ISReffect}. When integrated over all bins, the total relative bias on $\alpha_{\rm QED}(m^2_{\rm Z})$ amounts to $-8\times10^{-6}$, {\it i.e.}, well below the target statistical precision of a few $3 \times 10^{-5}$. The aforementioned theoretical knowledge of initial-state radiation and the in-situ determination of the angular resolution would allow this residual bias to be predicted and corrected for, with a precision at least an order of magnitude better. 

\subsubsection{Interference between initial- and final-state radiation (IFI)}

While initial-state radiation does not change the functional form of the muon angular distribution, the interference between initial-state and final-state occurs preferably when the final state muons are close to the initial state electrons, hence does affect their distribution in the forward and backward directions beyond the usual $(1+\cos^2 \theta^\ast) + 8/3 A_{\rm FB} \cos\theta^\ast$ formula. 

It is shown in Ref.~\cite{Abbiendi_2001} that the muon angular distribution is be modified by a multiplicative factor with a characteristic logarithmic dependence on $\cos\theta^\ast$, and can be parameterized as 
\begin{equation} 
\frac{{\rm d}\sigma_{\mu\mu}}{{\rm d}\cos\theta^\ast}(s^\prime) \propto
\left\{ 1+\cos^2 \theta^\ast + \frac{8}{3} A_{\rm FB}(s^\prime) \cos\theta^\ast \right\} \times
\left\{1 + f\left(\frac{s^\prime}{s}\right) \log\frac{1+\cos\theta^\ast}{1-\cos\theta^\ast}\right\}, 
\end{equation}
in presence of a tight muon acoplanarity cut as suggested in the previous section. The multiplicative factor contains an additional asymmetric term, which enhances the integrated muon forward-backward asymmetry. The tight cut on $s^\prime/s$ aimed at rejecting ISR also reduces IFI in similar proportions. To mitigate the very small residual effect of IFI on the angular distribution, the specific shape of the additional contribution can be fitted away, as was done at LEP and with the benefit of the much larger data samples expected at the FCC-ee. On the other hand, this additional contribution appears to be "universal", ({\it i.e.}, with an amplitude that depends only on $s^\prime/s$, similarly to what is observed for ISR), hence cancels out in the determination of $\alpha_{\rm QED}(m^2_{Z})$ from a combination of the measurements at the two centre-of-mass energies, with no loss of statistical power. 

\subsubsection{Other electroweak higher-order corrections}

Other electroweak corrections have so far "only" been computed off-peak with complete one-loop calculation~\cite{Bardin_2001}. One-loop box corrections lead to relative changes of $-9\times 10^{-4}$ at $\sqrt{s_-}$ and $-8 \times 10^{-4}$ at $\sqrt{s_+}$ from the improved Born approximation of $A_{\rm FB}^{\mu\mu} - A_{\rm FB,0}^{\mu\mu}$, hence to a shift of $\alpha_{\rm QED}(m_{\rm Z})$ at the per-mil level. A shift of similar size arises from one-loop vertex corrections. The theoretical uncertainty arising from the missing higher orders in the asymmetry calculation, estimated to be at the level of a few $10^{-4}$~\cite{Freitas_2015}, was adequate at the time of LEP but is insufficient today to match the precision offered by the FCC-ee. 

An order of magnitude improvement would be achievable today, with proven techniques, by including the dominant two-loop and leading three-loop corrections, and would represent a major breakthrough towards the FCC-ee targets. Meeting these targets might require a complete three-loop calculation, including three-loop box corrections, perhaps a serious challenge with the current techniques, and definitely beyond the scope of the present work. It is not unlikely, however, that a large part of these missing corrections affect in the same way the asymmetry at 87.9\,GeV and the asymmetry at 94.3\,GeV. If it were the case, the $\alpha_{\rm QED}(m^2_{Z})$ determination would enjoy a cancellation similar to the that observed for QED corrections, which could suffice even without a complete three-loop calculation. 

\section{Conclusions and outlook}
\label{sec:outlook}

In this paper, it has been shown that the measurement of the muon forward-backward asymmetry at the FCC-ee, with six months of data taking just below ($\sqrt{s} = 87.9$\,GeV) and just above ($\sqrt{s} = 94.3$\,GeV) the Z peak, as part of the Z resonance scan, would open the opportunity of a direct measurement of the electromagnetic constant $\alpha_{\rm QED}(m^2_{\rm Z})$, with a relative statistical uncertainty of the order of $3\times 10^{-5}$. 

A comprehensive list of sources for experimental, parametric, theoretical systematic uncertainties has been examined. Most of these uncertainties have been shown to be under control at the level of $10^{-5}$ or below, as summarized in Table~\ref{tab:summary}. A significant fraction of those benefits from a delicate cancellation between the two asymmetry measurements. The knowledge of the beam energy, both on- and off-peak, turns out to be the dominant contribution, albeit still well below the targeted statistical power of the method.   
\begin{table}[htbp]
\begin{center}
\begin{tabular}{|c|l|r|}
\hline Type & Source & Uncertainty \\ \hline\hline
                  & $E_{\rm beam}$ calibration & $1\times 10^{-5}$  \\
                  & $E_{\rm beam}$ spread      & $< 10^{-7}$  \\
Experimental      & Acceptance and efficiency  & negl.  \\
                  & Charge inversion & negl.  \\
                  & Backgrounds & negl. \\ \hline
                  & $m_{\rm Z}$ and $\Gamma_{\rm Z}$ & $1 \times 10^{-6}$ \\
Parametric        & $\sin^2\theta_{\rm W}$ & $5 \times 10^{-6}$ \\
                  & $G_{\rm F}$ & $5 \times 10^{-7}$ \\ \hline
                  & QED (ISR, FSR, IFI) & $< 10^{-6}$ \\ 
Theoretical       & Missing EW higher orders & few $10^{-4}$ \\ 
                  & New physics in the running &  $0.0$ \\ \hline\hline
Total            & Systematics & $1.2 \times 10^{-5}$ \\ 
(except missing EW higher orders)              & Statistics  & $3 \times 10^{-5}$ \\ \hline
\end{tabular}
\caption{Summary of relative statistical, experimental, parametric and theoretical uncertainties to the direct determination of the electromagnetic coupling constant at the FCC-ee, with a one-year running period equally shared between centre-of-mass energies of 87.9 and 94.3\,GeV, corresponding to an integrated luminosity of $85$\,${\rm ab}^{-1}$.}
\label{tab:summary}
\end{center}
\end{table}

The fantastic integrated luminosity and the unique beam-energy determination are {\it the} key breakthrough advantages of the FCC-ee in the perspective of a precise determination of the electromagnetic coupling constant. 
Today, the only obstacle towards this measurement -- beside the construction of the collider and the delivery of the target luminosities -- stems from the lack of higher orders in the determination of the electroweak corrections to the forward-backward asymmetry prediction in the standard model. With the full one-loop calculation presently available for these corrections, a relative uncertainty on $A_{\rm FB}^{\mu\mu}$ of the order of a few $10^{-4}$ is estimated. An improvement deemed adequate to match the FCC-ee experimental precision might require a calculation beyond two loops, which may be beyond the current state of the art, but is possibly within reach on the time scale required by the FCC-ee. 

A consistent international programme for present and future young theorists must therefore be set up towards significant precision improvements in the prediction of all electroweak precision observables, in order to reap the rewards potentially offered by the FCC-ee. 

\section*{Acknowledgements}

I thank Alain Blondel for enlightening discussions throughout the development of this analysis.
I am indebted to Roberto Tenchini for his expert suggestions and to Gigi Rolandi for his subtle comments and his consistent checking of all equations in the paper. I am grateful to Ayres Freitas for his careful reading, and for providing me with the predictions for $A_{\rm FB}^{\mu\mu}$ from {\tt ZFITTER} with the effects of the full one-loop calculation included.

\bibliography{AlphaEM.bbl}

\end{document}


\maketitle
\flushbottom

\section{Beam energy spread}
\label{sec:espread}

Equation~4.6 in Section 4.1.2 of Ref~\cite{Janot:2015gjr}, which gives the muon forward-backward asymmetry ($A_{\rm FB}^{\mu\mu}$) bias in presence of a finite beam energy spread $\delta$, is valid only under the assumption that the number of ${\rm e^+e^-} \to \mu^+\mu^-$ events is uniformly distributed over the centre-of-mass-energy interval covered by the beam energy spread. This assumption is not upheld by reality, as the dimuon production cross section $\sigma_{\mu\mu}$ follows the Z resonant lineshape, and therefore changes steeply with $\sqrt{s}$ around the two optimal centre-of-mass energies, $\sqrt{s_\pm}$. The equation should therefore read: 

\begin{equation}
\Delta A_{\rm FB}^{\mu\mu}(s_\pm) =  \frac{\int A_{\rm FB}^{\mu\mu}(s) \sigma_{\mu\mu}(s) \exp -\frac{\left( \sqrt{s}-\sqrt{s_\pm} \right)^2}{2s_\pm\delta^2} d\sqrt{s} }
{\int \sigma_{\mu\mu}(s) \exp -\frac{\left( \sqrt{s}-\sqrt{s_\pm} \right)^2}{2s_\pm\delta^2} d\sqrt{s}} - A_{\rm FB}^{\mu\mu}(s_\pm). 
\end{equation}

The effect of the forward-backward asymmetry reweighing by the production cross section amounts to increase (decrease) the average centre-of-mass energy above $\sqrt{s_-}$ (below $\sqrt{s_+}$), hence significantly change the average asymmetry towards smaller absolute values. Numerically, for a standard-model Z lineshape, and a relative beam energy spread of 0.12\%, the corresponding relative biases on the measured forward-backward asymmetries amount to 
\begin{equation} \frac{\Delta A_{\rm FB}}{A_{\rm FB}}(s_-) = -0.99 \times 10^{-3} {\rm \ \ \  and \ \ \ } \frac{\Delta A_{\rm FB}}{A_{\rm FB}}(s_+) = -1.03 \times 10^{-3},
\end{equation}
which differ from Eq.~4.8 of Ref.~\cite{Janot:2015gjr} both quantitatively: the effect is almost two orders of magnitude larger; and qualitatively: the previously claimed cancellation no longer occurs in the sum of Eq.~4.9. 
As a consequence, the bias on $\alpha_{\rm QED}$, which changes from $+1.6\times 10^{-6}$~\cite{Janot:2015gjr} to $-1.1 \times 10^{-3}$, needs to be corrected for, and therefore determined with a relative precision better than 1\%, to induce an uncertainty on $\alpha_{\rm QED}$ of the same order as or smaller than that arising from the beam energy measurement. 

A method to monitor the centre-of-mass energy spread with this precision every couple minutes at the FCC-ee, from the angular analysis of the same ${\rm e^+e^-} \to \mu^+\mu^-$ events as those used for the forward-backward asymmetry measurement, is written up in  the forthcoming FCC Conceptual Design Report~\cite{FCCeeCDR}. In Table~1 of Ref.~\cite{Janot:2015gjr}, the line "$E_{\rm beam}$ spread" is consequently modified from $< 10^{-7}$ to $ < 10^{-5}$. 

The conclusions of Ref.~\cite{Janot:2015gjr} are unchanged after these corrections. 

\section{Interference between Initial and Final state radiation (IFI) }

In Section~4.3.3 of Ref.~\cite{Janot:2015gjr}, it is stated that {\it the tight cut on $s^\prime/s$ aimed at rejecting ISR also reduces IFI in similar proportions.} This statement is incomplete. While rejecting indeed events with energetic photons in similar proportions with or without IFI, a tight cut on $s^\prime/s$ has no effect on soft photons, which in turn have a macroscopic effect (several per-cent increase) on the numerical value of the muon forward-backward asymmetry when IFI is included. As pointed out in Ref.~\cite{JadachFCCWeek}, an important cancellation, already exploited in Ref.~\cite{Janot:2015gjr} for ISR, still occurs for IFI between the two centre-of-mass energies $\sqrt{s_-}$ and $\sqrt{s_+}$. 

The precision with which this cancellation can be predicted is being evaluated and is the topic of two forthcoming publications~\cite{Jadach:2018a, Jadach:2018b}.
\acknowledgments

I would like to thank very much Mogens Dam for kindly pointing out that Eq.~4.6 of Ref.~\cite{Janot:2015gjr} was missing an important contribution, and for motivating me to ascertain the robustness of the centre-of-mass energy spread determination with a sub-per-cent precision at the FCC-ee. 

\bibliographystyle{JHEP}
\bibliography{biblio.bib}